\documentclass[article]{jss} 
\usepackage[utf8]{inputenc}
\usepackage{color}

\usepackage[british]{babel} 
\author{Przemys\l{}aw Biecek\\University of Warsaw \And 
        Marcin Kosi\'nski\\Warsaw University of Technology}
\title{\pkg{archivist}: An \proglang{R} Package for Managing, Recording and Restoring Data Analysis Results}

\Plainauthor{Przemyslaw Biecek, Marcin Kosinski} 
\Plaintitle{archivist: An R Package for Managing, Recording and Restoring Data Analysis Results} 
\Shorttitle{R Package \pkg{archivist}} 

\Abstract{
Everything that exists in \proglang{R} is an object \cite{Chambers2016}. This article examines what would be possible if we kept copies of all \proglang{R} objects that have ever been created. Not only objects but also their properties, meta-data, relations with other objects and information about context in which they were created.

We introduce \pkg{archivist}, an \proglang{R} package designed to improve the management of results of data analysis. Key functionalities of this package include: (i) management of local and {remote} repositories which contain \proglang{R} objects and their meta-data (objects' properties and relations between them); (ii) {archiving} \proglang{R} objects to repositories; (iii) sharing and retrieving objects (and it's pedigree) by their unique hooks; (iv) searching for objects with specific properties or relations to other objects; {(v) verification of object's identity and context of it's creation.}

The presented \pkg{archivist} package extends, in a combination with packages such as \pkg{knitr} {and} \pkg{Sweave}, the reproducible research paradigm by creating  new ways to retrieve and validate previously calculated objects. These new features give a variety of opportunities such as: sharing \proglang{R} objects within reports or articles; adding hooks to \proglang{R} objects in table or figure captions; interactive exploration of object repositories; caching function calls with their results; retrieving object's pedigree (information about how the object was created); automated tracking of the performance of considered models, restoring \proglang{R} libraries to the state in which object was archived.
}
\Keywords{recordable research, reproducible research, data analysis management, data governance, meta-data, results management}

\Address{
Przemys\l{}aw Biecek\\
Faculty of Mathematics, Informatics, and Mechanics\\ 
University of Warsaw\\
Banacha 2, 02-097 Warsaw, Poland \\
E-mail: \email{Przemyslaw.Biecek@gmail.com}
\\
\
\\
Marcin Kosi\'nski\\
Faculty of Mathematics and Information Science\\ 
Warsaw University of Technology\\
Koszykowa 75, 00-662 Warsaw, Poland \\
E-mail: \email{M.P.Kosinski@gmail.com}
}
    
\begin{document}


\section{Introduction}

In most of the cases the outcome of the process of data analysis is a set of objects in the form of statistical models, charts or tables. Three requirements are often superimposed to ensure sufficient quality of such results: they should be reproducible, verifiable and accessible. Reproducibility means that there is a process that reproduces results. Verifiability means that it is possible to check whether the newly generated results are identical to previously obtained results{, and it is possible to check the context of object's creation}. Accessibility means that results can be easily accessed for future computer based processing. Reproducibility gets increasing attention in the academic literature across various disciplines, see for example \cite{Peng01072009} for bioinformatics or \cite{Koenker} for the econometric research or \cite{Replicability} for more general discussion about differences between replicability and reproducibility. 

The \proglang{R} ecosystem of packages is equipped with wonderful tools such as \pkg{knitr} \citep[see][]{knitr2, knitr} or \pkg{Sweave} \citep[see][]{Sweave, Sweave2} which allow to create reproducible reports or articles. They follow the \textit{literate programming} principle, and the R code, its results and its explanations appear together in a single document. {It is assumed that} the same input and identical instructions executed on the same operating system with the same local settings and with identical versions of installed libraries will result in the same output. Under these assumptions \pkg{knitr} or \pkg{Sweave} reports are sufficient to recreate the previously obtained results. 

But there are cases in which it is not convenient to recreate results from scratch, from raw input. Consider the following situations: 
\begin{itemize}
\item the input data is large or with limited/restricted access (e.g., for genomic data the raw input may easily hit few TB); 
\item computations take a lot of time or require specialized hardware (e.g., calculations tuned for Graphics Processing Unit cards); 
\item calculations are based on a very specific version of software or require commercial versions of software or some functions are  deprecated or removed over time. It can be an issue even for open software, e.g., due to rapid development of \proglang{R}, even widely used packages experience significant changes, like \pkg{ggplot2} or \pkg{lme4} in the year 2015;
\item results are generated and processed periodically and you wish to restore and compare models across all reports. 
\end{itemize}

In such situations it is desirable to retrieve the results that were calculated in the past rather than reproducing them from scratch. Objects that are backed up can be reused even if they cannot be reproduced or the reproducibility will be too complex or time consuming. Alternatively, it may be desired to check whether the reproduced results are the same as those obtained previously.

An interesting example of such a feature are StatLinks \citep[see][]{statlink} commonly used in reports prepared by OECD (Organization for Economic Co-operation and Development). In addition to scripts that generate results, most tables and plots that are presented in the reports are equipped with their own DOIs (digital object identifier) and web hooks. Through these links readers may download selected tables and plots, in the Excel format. 
The \code{xls} and \code{xlsx} formats are not ideal as they are proprietary and difficult to read in an automated way. But for extensive studies it is convenient and faster to access final results in such formats instead of having scripts that reproduce them.

If the only result from the data analysis is a single plot, a model or a table, it is easy to save it in the \code{rda} format and make it accessible for the others. But increasing amounts of heterogeneous data results in growing complexity of the process of data analysis. The complexity comes either from data volume, data heterogeneity, numerous steps required for data preparation, results validation etc. Moreover, working with data is often a highly iterative process that generates large amount of partial or final results. For all the above reasons the management of versions of results becomes a task in itself. Neglecting this process results in \textit{Reproducibility Debt} and may consequently lead to huge additional workload when it comes to recreation of results. The \textit{Reproducibility Debt} is a part of wider category called \textit{Technical Debt} \citep[see][]{TechnicalDebt}.

It should be noted that the concept of recording and exploring relations between objects is not new. Potential applications in auditable data analyses were  discussed almost 30 years ago \citep[see][]{Auditing}. What we present in this article may be perceived as implementation of some of these concepts. It is now easier due to lower costs of data storage.

The \pkg{archivist} package helps in managing, sharing, storing, linking and searching for \proglang{R} objects {in a platform agnostic way}. Its core functionalities allow for many interesting applications - some of them are presented in the Section \ref{sec:mot2}. The \pkg{archivist} package automatically retrieves the object's meta-data and creates a rich structure that allows for easy management of stored \proglang{R} objects. The meta-data covers object's properties such as: name, creation date, class, versions of attached packages, structure and relations between \proglang{R} objects (as for example, that an object \textit{A} was used for creation of an object \textit{B}). All examples presented here are related to \proglang{R} objects. In the Section \ref{sec:extensions} we discuss how this approach can be extended to other languages.

The rest of the article has the following structure. In the Section \ref{sec:mot2} (\textit{Motivation}) we introduce key motivations and use-cases behind \pkg{archivist}. In the Section \ref{sec:func3} (\textit{Functionality}) we present all functions available in the package and point out some further directions how this functionality can be integrated with GitHub, \pkg{knitr}, or be extended on other languages / formats. In the Section \ref{sec:conc4} (\textit{Conclusions}) we gather some final thoughts related to recordable and restorable research.

\section{Motivation}
\label{sec:mot2}

In this section we present key concepts and some use-cases behind the \pkg{archivist} package. In the Section \ref{sec:func3} we present all functions available in the \pkg{archivist} in a more formal way. First let us introduce some terminology. 

\begin{itemize}
\item Artifact - an \proglang{R} object that is saved to the repository. Artifacts are identified by their MD5 hashes.
\item Repository - a collection of artifacts stored as binary files outside of the \proglang{R} session. Repositories are either local (with a read-write access) or {remote} (with a read access only). The API for repositories allow for following actions: add, delete, read or search for an artifact with selected Tags. In the current version of the \pkg{archivist} local repositories are folders in the file system while remote repositories are Git or Mercurial based repositories. The same mechanism can be used to access repositories pointed as URL addresses or folders attached to \proglang{R} packages.
\item MD5 hash - a unique identifier of an artifact. It's a 32-character-long string, result of cryptographical hash function MD5 (Message Digest algorithm 5). Here, we are using implementation of hash function available in the  \pkg{digest} package \citep[see][]{digest}. In the \pkg{archivist} package MD5 hashes are used as object's hooks. 
\item Tag  - an attribute of an artifact. Tags are represented as character strings; they usually have the following structure: \code{key:value}. An artifact may have many tags, even with the same key. Some tags are automatically derived from artifacts, others may be added manually. Tags may be referred as meta-data of artifacts as they describe either properties of artifacts (e.g., class, name, date of creation) or relations between artifacts (e.g., being a part of, being a result of).
\end{itemize}

The \pkg{archivist} package manages \proglang{R} objects outside the \proglang{R} session. It stores binary copies of \proglang{R} objects in \code{rda} files and provides easy access for seeking and restoring these objects based on timestamps, classes or other properties.

\begin{figure}[b!h]
\centering
\includegraphics[width=0.8\textwidth]{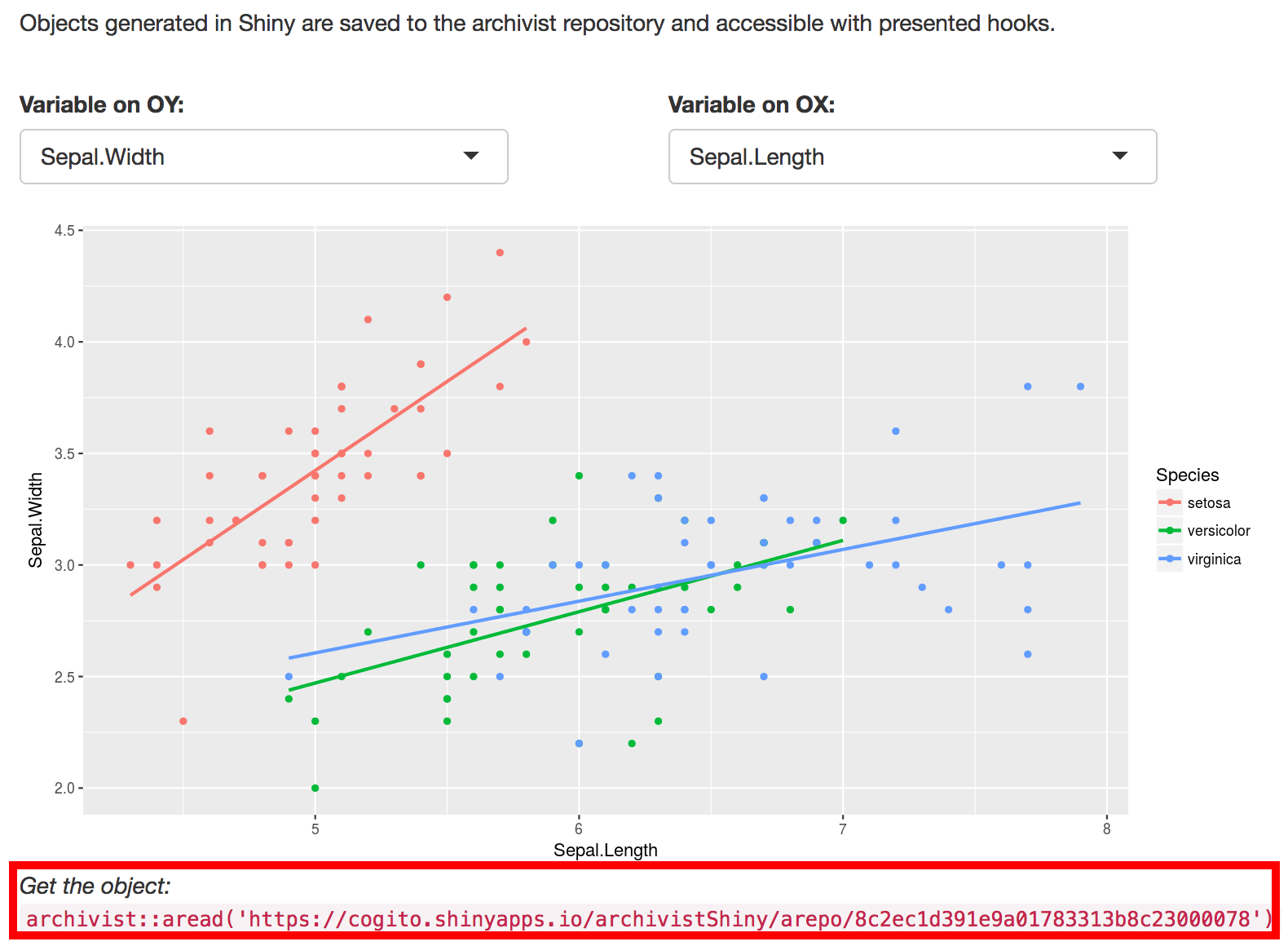}
\caption{\label{fig:shinyArchivist}A screenshot from a Shiny application hosted under the link \texttt{https://cogito.shinyapps.io/archivistShiny}. The \pkg{archivist} hook is included below each plot.}
\end{figure}

But, why anybody would like to store copies of \proglang{R} objects? Let's imagine the following use-cases:

\begin{itemize}
\item A data {scientist} creates a report or an article and would like to provide an access to results presented in the article. Typically, these results are presented as plots, tables or models. Apart from including these results in the report or article in a human-readable form, it may be beneficial to be able to restore a given result in a machine-readable form for further processing. Having a possibility to retrieve an \proglang{R}  plot or table, one can perform some further transformation of it. The opportunity to retrieve a regression model enables additional residuals' validation or applying model to the new data. The \pkg{archivist} creates a hook to a copy of \proglang{R} object which restores the object in a remote \proglang{R} session. Such hooks are short one-line instructions and can be embedded in figures' or tables' captions. 

An example report that illustrates this use-case is available at \texttt{http://bit.ly/1nW9Cvz}. A part of it is presented in Figure \ref{figExample3}. The report is created with the use of \pkg{knitr} package. It contains both \proglang{R} code and it's results in the form of tables and plots created with \pkg{ggplot2} package \citep[see][]{ggplot2}. In addition, there are also hooks to selected results. These hooks allow to restore a given plot or table directly in the local R session. Hooks of such a form restore a \code{gg} object in an R session.

\code{archivist::aread("pbiecek/Eseje/arepo/ba7f58fafe7373420e3ddce039558140")}

\begin{figure}[b!h]
\centering
\includegraphics[width=0.9\textwidth]{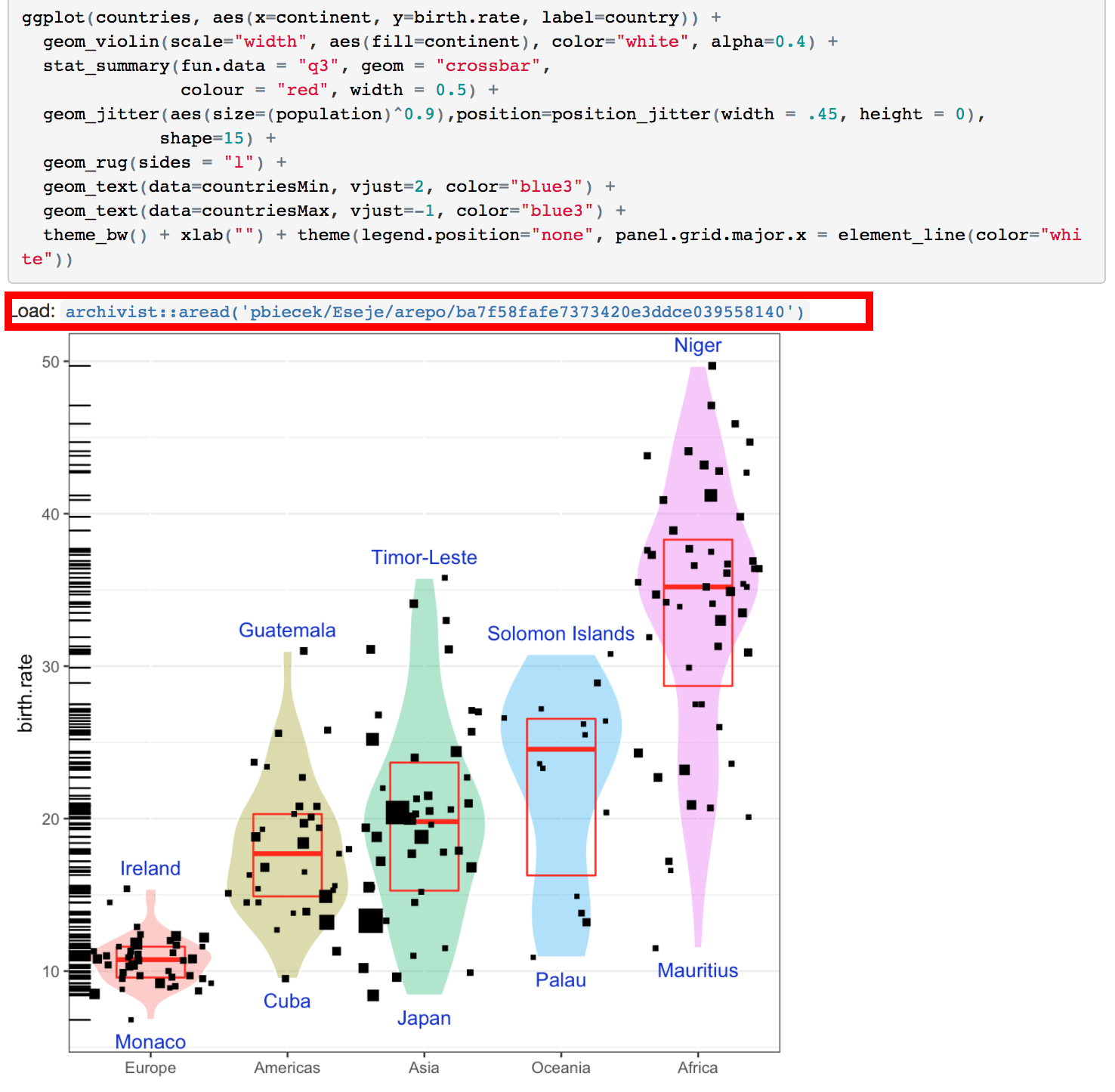}
\caption{\label{figExample3}{A part of a \pkg{knitr} report \texttt{http://bit.ly/1nW9Cvz} that uses the \code{addHooksToPrint} function that automatically adds \pkg{archivist} hooks to all objects of a given class. Objects can be accessed either by copying highlighted \code{aread} instructions to \proglang{R} or by clicking the link.}}
\end{figure}

\item A team of data scientists is working for some time on a forecasting model. During a certain period of time a large set of competing models is created. The team needs a tool that stores all models with additional metadata, such as model performance, information which data was used for model training and testing. The \pkg{archivist} creates a shared repository which can be used for storing models along with their meta-data and provides API for searching objects with specific meta-data. The example below reads all objects of the class \code{lm}, calculates a BIC score for them and sorts objects with respect to these scores.

\begin{Sinput}
R> library("archivist")
R> models <- asearch("pbiecek/graphGallery", patterns = "class:lm")
R> modelsBIC <- sapply(models, BIC)
R> sort(modelsBIC)
\end{Sinput}

\begin{Soutput}
990861c7c27812ee959f10e5f76fe2c3 2a6e492cb6982f230e48cf46023e2e4f 
                        39.05577                         67.52735 
0a82efeb8250a47718cea9d7f64e5ae7 378237103bb60c58600fe69bed6c7f11 
                       189.73593                        189.73593 
7f11e03539d48d35f7e7fe7780527ba7 c1b1ef7bcddefb181f79176015bc3931 
                       189.73593                        189.73593 
0e213ac68a45b6cd454d06b91f991bc7 e58d2f9d50b67ce4d397bf015ec1259c 
                       243.49450                        243.49450 
18a98048f0584469483afb65294ce3ed 
                       396.16690
\end{Soutput}

\item Results are generated in a remote \proglang{R} process, like for example with a Shiny application. The \pkg{archivist} saves created \proglang{R} artifacts in an URL repository. 

See for example Figure \ref{fig:shinyArchivist} that presents a screenshot from the Shiny application \linebreak\texttt{https://cogito.shinyapps.io/archivistShiny}.
All plots generated by this application are stored in an \pkg{archivist} repository and may be accessed with hooks presented below plots. Following line downloads a single plot directly to the local \proglang{R} session.

\code{archivist::aread("https://cogito.shinyapps.io/archivistShiny/arepo/}

\code{ca680b829abd8f0a4bd2347dcf9fe534")}.
\end{itemize}

\begin{figure}[h!]
\centering
\includegraphics[width=0.9\textwidth]{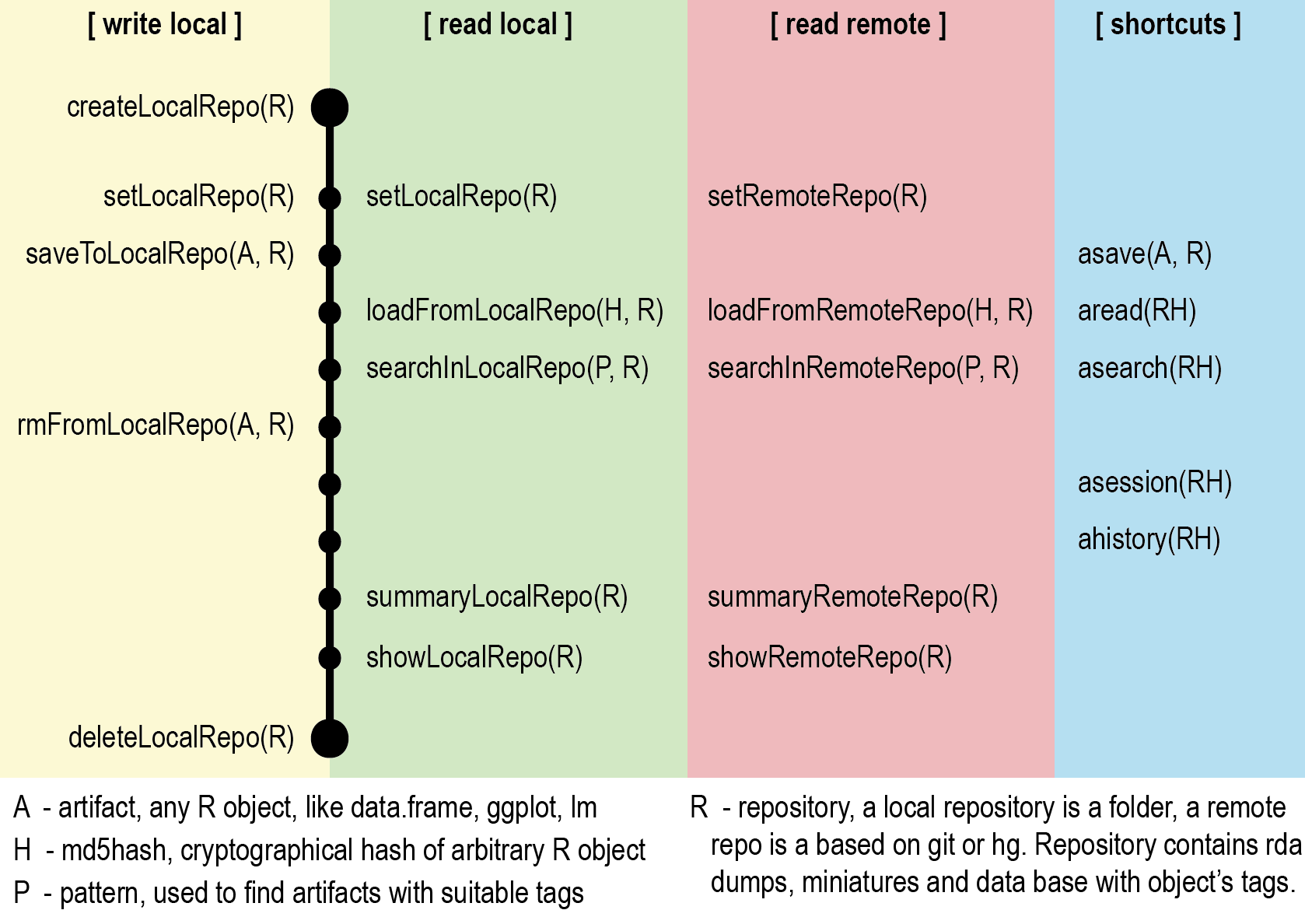}
\caption{\label{fig:structure}Overview of the most important functions related to a life-cycle of a repository and an artifact.}
\end{figure}

\section{Functionality}
\label{sec:func3}

The key functionality of the \pkg{archivist} package is to manage copies of \proglang{R} objects, called artifacts, stored as binary files. Artifacts are stored in collections called repositories. Properties of artifacts and relations between artifacts are described by their tags.

Typical lifetime of the repository is presented in Figure \ref{fig:structure}. The local repository is created with the \code{createLocalRepo} function. It can be set as a default repository so that calls of the other \pkg{archivist} functions can be simplified. 
Once the repository is created, new \proglang{R} objects can be archived with the \code{saveToLocalRepo} function or can be removed with the \code{rmFromLocalRepo} function. Artifacts can be restored from the repository with \code{loadFromLocalRepo} function. One can also get all objects that match given criteria with the function named \code{searchInLocalRepo}. Both functions have wrappers called \code{aread} and \code{asearch}, respectively, with the simplified and shorter interface.
To summarise what kind of artifacts are in the repository one can use \code{summaryLocalRepo} or \code{showLocalRepo} functions. The repository can be removed with the \code{deleteLocalRepo} function.

Table \ref{tab:functions} presents all functions available in the \pkg{archivist} package. These functions are divided into four core groups:

\begin{itemize}
\item Functions for repository management. In this group there are functions used to create a new empty repository, to create a repository as a copy of an existing local or GitHub repository, to backup an entire repository into a single \textit{zip} file, to present summary statistics of objects stored in the repository and to delete existing repository.
\item Functions for saving artifacts to a repository, loading artifacts from a repository and removing artifacts from a repository. Functions that show relations between artifacts, present artifacts' history or context in which they were created.
\item Functions for searching for artifacts within a repository. Artifacts may be accessed through date of creation, a tag or a list of tags.
\item Other features that do not fit previous categories.
\end{itemize}

In sections \ref{sec:repo31}-\ref{sec:extensions} each group of these functions is presented separately.

\begin{table}[t!]
\begin{center}
\begin{tabular}{lll}
\hline
	&	\textbf{Local}	&	\textbf{Remote}	\\ \hline
\textit{Repository managment}	&	createLocalRepo	&		\\
	&	setLocalRepo	&	setRemoteRepo	\\
	&	deleteLocalRepo	&		\\ 
	&	showLocalRepo	&	showRemoteRepo	\\
	&	summaryLocalRepo	&	summaryRemoteRepo	\\  
	&	zipLocalRepo	&	zipRemoteRepo	\\ 
	&	copyLocalRepo	&	copyRemoteRepo	\\ \hline
\textit{Artifacts management} 	&	saveToLocalRepo	&		\\
	&	rmFromLocalRepo 	&		\\  
	&	loadFromLocalRepo	&	loadFromRemoteRepo	\\
	&	aread &	aread	\\
	&	asession &	asession	\\
	&	aformat &	aformat	\\
	&	ahistory	 &		\\ 
	&	\code{\%a\%}	 &		\\  \hline
\textit{Artifacts' exploration}	&	searchInLocalRepo	&	searchInRemoteRepo \\
	&	asearch	&	asearch	\\ 
	&	shinySearchInLocalRepo	&	 	\\ \hline
\textit{Extensions}	&	restoreLibs 	&	 \\
	&	atrace 	&		\\ 
	&	addHooksToPrint	&		\\ 
	&	createMDGallery	&		\\ \hline
\end{tabular}
\caption{\label{tab:functions} The list of functions available in the \pkg{archivist} package version 2.1. }
\end{center}
\end{table}

\subsection{Repository management}
\label{sec:repo31}

A repository is a collection of artifacts and their meta-data. In this section you will find a~list of functions for repository management (used to create a~new empty repository, create a copy, present summary statistics or delete existing repository).

\begin{figure}[b!]
\centering
\includegraphics[width=0.6\textwidth]{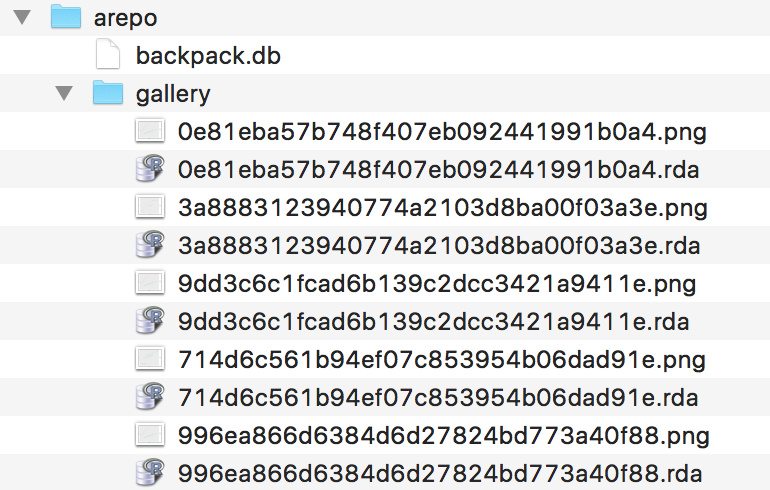}
\caption{\label{fig:dirb}The structure of an example \code{arepo} repository. It contains database with objects' meta-data stored in an SQLite file \code{backpack.db} and a subfolder \code{gallery} with binary copies of \proglang{R} objects and their miniatures.}

\centering
\includegraphics[width=0.8\textwidth]{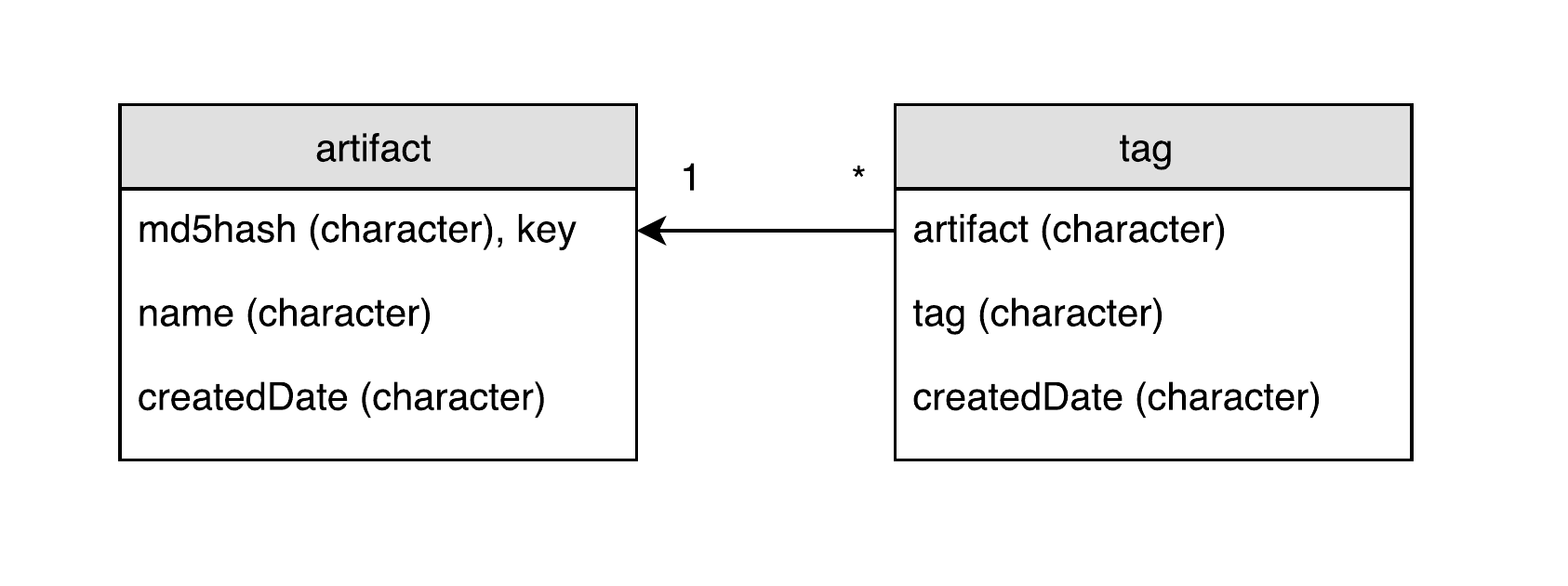}
\caption{\label{ERD}The Entity-Relationship Diagram that presents the structure of tables: \textit{artifact} and \textit{tag}, summarizing the relations between artifacts. The SQLite database with both tables is stored in \code{backback.db} file in the repository.}
\end{figure}

Technically, repository is a directory with the following structure (see Figure \ref{fig:dirb}).
\begin{itemize}
\item A \code{backpack.db} file which contains an SQLite database. The database contains two tables with a structure presented in Figure \ref{ERD}. The table named \textit{artifact} contains artifacts’ MD5 hashes and basic information about the artifacts. The table called \textit{tag} contains artifacts’ tags. Since both artifacts and tags may be added into the database an unspecified number of times, each tag and artifact has one or more time points - one for each attempt to artifact's or tag's archiving to the repository.
\item A subdirectory called \code{gallery} with artifacts' storage. Artifacts are stored as separate files.  {Names of files start with MD5 hashes of corresponding artifacts. Extensions correspond to formats in which artifacts are saved. The current implementation for \proglang{R} stores artifacts in the \code{rda} format, but it can be easily extended to handle other formats. Additionally, also an artifact’s miniature is saved.} 
For plots the default format for miniatures is raster file with  \code{png} extension, for other objects it is a text file with \code{txt} extension (e.g., for data frames it contains first few rows).
\end{itemize}

A repository may be accessed in two ways.
\begin{itemize}
\item Local - in this case repository is identified by its path in the local file system. The repository is in the read-write mode. If the file system is shared (shared file system on HPC cluster, a Dropbox directory, a mounted folder on Network File System, Secure Shell Filesystem, etc.) multiple users may read and write into the repository at the same time. 
\item { Remote} - {Currently \pkg{archivist} supports GitHub and Bitbucket repositories, but it can be easily extended to support any git or mercurial repository, see Section 4.} Repository is identified by it's type (github/bitbucket), a username and the repository's name. The repository is accessible in read-only mode. Multiple users can read from such repository at the same time. 
In order to write to a remote repository one should either synchronize a~local directory with GitHub/Bitbucket account or use a \pkg{archivist.github} package, which is \pkg{archivist}'s first extension \citep[see][]{archivist.github}.
\end{itemize}

\clearpage

The logic behind this is as follows. 
Depending on the user’s needs it is possible to create a~single repository per project or per group of projects or keep all artifacts ever created in a~single repository. Since (i) a local repository is accessible even without an Internet connection, (ii) the access is faster and (iii) there is both read and write access, it is easier to work with local repositories, which are just a directory identified by its path. 
If the user wants to share a~repository with artifacts with a general public then he or she can publish the local repository on GitHub {or Bitbucket} or make it available as a subdirectory of an \proglang{R} package. 

\subsubsection{Creation of a new empty repository}

The \code{createLocalRepo} function creates a new local repository. The \code{repoDir} argument points to a directory that will be used as a repository root. The directory will be created if it does not exist. The \code{default=TRUE} argument marks the newly created repository as a default one.

The directory may be specified either by global path or local path. {The example below will create a repository named \code{arepo} in the current working directory.}

\begin{Sinput}
R> repo <- "arepo"
R> createLocalRepo(repoDir = repo, default = TRUE)
\end{Sinput}

\subsubsection{Deletion of an existing repository}

The \code{deleteLocalRepo} function deletes all artifacts, miniatures, the database with meta-data and the directory identified by the \code{repoDir} argument.

\begin{Sinput}
R> repo <- "arepo"
R> deleteLocalRepo(repoDir = repo)
\end{Sinput}

\subsubsection{Copying artifacts from other repositories}

Functions \code{copyLocalRepo} and \code{copyRemoteRepo} copy selected artifacts from either local or remote (GitHub or Bitbucket) repository into a local repository. Artifacts to be copied are identified by their MD5 hashes. 

In the example below the artifact identified by hash {\code{7f3453331910e3f321ef97d87adb5bad}} is copied along with its meta-data from remote GitHub repository  \code{pbiecek/graphGallery} to the local repository \code{arepo}.

\begin{Sinput}
R> repo <- "arepo"
R> createLocalRepo(repoDir = repo, default = TRUE)
R> copyRemoteRepo(repoTo = repo, 
+    md5hashes = "7f3453331910e3f321ef97d87adb5bad",
+    user = "pbiecek", repo = "graphGallery", repoType = "github")
\end{Sinput}

Functions \code{zipLocalRepo} and \code{zipRemoteRepo} download all artifacts and create a single \code{zip} archive.

\subsubsection{Showing repository's statistics}

A repository is a collection of artifacts and their meta-data. Functions \code{summaryLocalRepo} and \code{summaryRemoteRepo} summarize basic statistics about artifacts in the repository. Functions \code{showLocalRepo} and \code{showRemoteRepo} list all MD5 hashes and artifact’s meta-data. 

Functions \code{show*Repo} take argument \code{method} which may be either \code{"tags"} (the result is a data frame with artifact’s tags) or \code{"md5hashes"} (default, result is a data frame with artifact’s MD5 hashes). 

In the previous example we copied a single artifact from GitHub repository to the local one. The artifact is copied with its tags. In the example below we list all the tags within this single-artifact repository.

\begin{Sinput}
R> showLocalRepo(repoDir = repo, method = "tags")
\end{Sinput}
\begin{Soutput}
                        artifact                      tag         createdDate
1 7f3453331910e3f321ef97d87adb5b               format:rda 2016-02-09 14:37:06
2 7f3453331910e3f321ef97d87adb5b                 class:gg 2016-02-09 14:37:06
3 7f3453331910e3f321ef97d87adb5b             class:ggplot 2016-02-09 14:37:06
4 7f3453331910e3f321ef97d87adb5b      labelx:Sepal.Length 2016-02-09 14:37:06
5 7f3453331910e3f321ef97d87adb5b      labely:Petal.Length 2016-02-09 14:37:06
6 7f3453331910e3f321ef97d87adb5b date:2016-02-09 14:37:06 2016-02-09 14:37:06
7 7f3453331910e3f321ef97d87adb5b               format:png 2016-02-09 14:37:06
8 ff575c261c949d073b2895b05d1097 relationWith:2166d...... 2015-06-22 17:17:14
8 ff575c261c949d073b2895b05d1097 sessionInfo:3b8c60...... 2015-06-22 17:17:14
\end{Soutput}

In the example below the function \code{summaryLocalRepo} is used to list summaries of artifacts in the repository called \code{graphGallery} which is attached to the \pkg{archivist} package. One can find information about dates on which artifacts were added, classes of artifacts and the total number of artifacts in the repository.

\begin{Sinput}
R> summaryLocalRepo(repoDir = 
+    system.file("graphGallery", package = "archivist")) 
\end{Sinput}
\begin{Soutput}
Number of archived artifacts in Repository:  7 
Number of archived datasets in Repository:  3 
Number of various classes archived in Repository: 
            Number
gg              2
ggplot          2
lm              3
data.frame      2
summary.lm      1
Saves per day in Repository: 
            Saves
2016-02-07     6
2016-02-08    15
\end{Soutput}

\subsubsection{Setting a default repository}

In most of the cases we work with one repository per project. In such cases it is convenient to set a default local or remote repository. It can be done with \code{setLocalRepo} or \code{setRemoteRepo} functions. Look at the example below.

\begin{Sinput}
R> setRemoteRepo(user = "pbiecek", repo = "graphGallery", repoType = "github")
R> setLocalRepo(repoDir = 
+    system.file("graphGallery", package = "archivist"))
\end{Sinput}

After setting a default repository, one can use the following functions
\begin{itemize}
\item \code{saveToLocalRepo},
\item \code{loadFromLocalRepo}, \code{loadFromRemoteRepo}, 
\item \code{rmFromLocalRepo}, 
\item \code{searchInLocalRepo}, \code{searchInRemoteRepo}, 
\end{itemize}
without specification of \code{repoDir} or \code{user/repo/branch/subdir/repoType} arguments. 

For example, the instruction below will add \code{iris} data frame to the default local repository.

\begin{Sinput}
R> setLocalRepo(repoDir = repo)
R> data("iris")
R> saveToRepo(iris)
\end{Sinput}

Another option for setting a default value for an argument is the function \code{aoptions()}. It sets the default value for any argument that is used by \code{archivist}. For example the instruction below sets the default value for \code{repoType} to \code{"github"}.
\begin{Sinput}
R> aoptions("repoType", "github")
\end{Sinput}


\subsection{Artifact management}
\label{sec:art32}

An artifact is an \proglang{R} object with its meta-data. Artifacts are stored in repositories. Key functions for artifact's management are functions for saving, loading and removing artifacts from a repository. 

\subsubsection[Saving an R object into a repository]{Saving an \proglang{R} object into a repository}

The \code{saveToLocalRepo} function saves any \proglang{R} object into the selected repository. It stores in the repository both the object and its tags. 
Some tags and some {meta-data} are extracted in an automated way. The \code{saveToLocalRepo} function recognizes the class of the artifact and extracts tags typical for that class. It is possible to add support for a new class of objects or change list of tags extracted for selected classes, just extend the generic function \code{extractTags()}. Table \ref{tab:tags} lists classes that are recognized in the current version of the package and lists tags that are derived automatically from objects of a given class. For other classes the following attributes are extracted: name, creation time and MD5 hash.

\begin{table}[h!]
\begin{center}
\begin{tabular}{p{3cm}p{11cm}}
\hline
Artifact's class & Tags \\ \hline
lm & date, name, class, coefname, rank, df.residual  \\ 
survfit & date, name, class, strata, type, n, conf.type, conf.int \\ 
ggplot & date, name, class, labelx, labely \\ 
twins & date, name, class, ac \\ 
partition & date, name, class, objective, memb.exp, coeff, k.crisp, conv, clus.avg.widths, avg.width \\ 
qda & date, name, class, terms, N, lev, counts, prior, ldet \\ 
lda & date, name, class, N, lev, counts, prior, svd \\
htest & date, name, class, alternative, method, data.name, null.value, statistic, parameter, p.value, intervals, estimate\\ 
data.frame & date, name, class, varname \\ 
summary.lm & date, name, class, sigma, df, r.squared, adj.r.squared, fstatistic, fstatistic.df \\
glmnet & date, name, class, dim, nulldev, npasses, offset, nobs \\
default & date, name, class \\ 
\hline
\end{tabular}
\caption{\label{tab:tags}Tags that are automatically extracted from objects depending on the object's class. See \code{?Tags} for more details.}
\end{center}
\end{table}

The \code{saveToLocalRepo} function takes at least two arguments: \code{artifact} - an \proglang{R} object which is about to be saved and \code{repoDir} which is a path to the local repository.
The process of adding an \proglang{R} object to the repository triggers a chain of actions listed below. By setting some arguments of \code{saveToLocalRepo} to \code{FALSE} some of these actions may be skipped.
\begin{itemize}
\item {The} name of the object is derived and stored as the object's tag \code{name:xxx}. It may be useful when searching for an object. One can search for all objects that had a specific name with \code{asearch(pattern="name:iris")}. 
\item An MD5 hash is calculated for the object {with the use of \pkg{digest} package.} Then the object is saved as a binary file named \code{md5hash.rda} with the use of \code{save} function. 
\item If there is any dependent object, it is saved separately to the repository (e.g., for object of class \code{gg} or \code{lm} the \code{data} slot is extracted from the object and saved separately. Additionally a tag \code{relationWith:xxx} is added, where \code{xxx} is the MD5 hash of the dataset).
\item The current session info, with the list of versions of attached packages, is saved to the repository.  The session info is linked to the artifact. The link is a tag of the form \code{sessionInfo:xxx}, where \code{xxx} stands for MD5 hash of the object with session info. 
\item A set of tags is extracted automatically and these tags are saved to the repository. See Table \ref{tab:tags} for the list of tags that are automatically derived. Tags extracted for a given class are defined by the generic \code{extractTags} function.
\item Additional tags specified by a user (with the \code{userTags} argument) are saved to the repository as well.
\item A miniature for the object is created – for plots it is a png file while for data frames or models it is a text description of the object.
\end{itemize}

The following example creates a plot of the class \code{gg} and saves the object into the repository. Plots created with the use of \pkg{ggplot2} package are objects and can be serialized in the same way as any other \proglang{R} objects \citep[see][]{ggplot2}. A hash of the recorded object is returned. In the example below it is \code{11127cc6ce69a89d11d0e30865a33c13}. By default, the related data object is also saved. In this case the dependent object is a dataset \code{iris} which is saved with the hash \code{ff575c261c949d073b2895b05d1097c3}.
\begin{Sinput}
R> library("ggplot2")
R> repo <- "arepo"
R> pl <- qplot(Sepal.Length, Petal.Length, data = iris)
R> saveToLocalRepo(pl, repoDir = repo)
\end{Sinput}
\begin{Soutput}
[1] "11127cc6ce69a89d11d0e30865a33c13"
attr(,"data")
[1] "ff575c261c949d073b2895b05d1097c3"
\end{Soutput}

The function \code{saveToLocalRepo} extracts additional tags such as the name of the original object (here: \code{name:pl}), its class (\code{class:gg}), labels on OX and OY axes (\code{labelx:Sepal.Length}) and MD5 hash of the data object. These tags are listed if we use \code{showLocalRepo} function on the repository.

\begin{Sinput}
R> showLocalRepo(repoDir = repo, "tags")
\end{Sinput}
\begin{Soutput}
                         artifact                     tag         createdDate
1  11127cc6ce69a89d11d0e30865a33c              format:rda 2016-02-09 16:42:59
2  11127cc6ce69a89d11d0e30865a33c                 name:pl 2016-02-09 16:42:59
3  11127cc6ce69a89d11d0e30865a33c                class:gg 2016-02-09 16:42:59
4  11127cc6ce69a89d11d0e30865a33c            class:ggplot 2016-02-09 16:42:59
5  11127cc6ce69a89d11d0e30865a33c     labelx:Sepal.Length 2016-02-09 16:42:59
6  11127cc6ce69a89d11d0e30865a33c     labely:Petal.Length 2016-02-09 16:42:59
7  11127cc6ce69a89d11d0e30865a33c  date:20160209 16:42:59 2016-02-09 16:42:59
8  11127cc6ce69a89d11d0e30865a33c session_info:e0373..... 2016-02-09 16:42:59
9  11127cc6ce69a89d11d0e30865a33c              format:png 2016-02-09 16:42:59
10 e037375a5f757efcc28561c0a1a2ef              format:rda 2016-02-09 16:42:59
11 ff575c261c949d073b2895b05d1097              format:rda 2016-02-09 16:42:59
12 ff575c261c949d073b2895b05d1097 session_info:e0373..... 2016-02-09 16:42:59
13 ff575c261c949d073b2895b05d1097              format:txt 2016-02-09 16:42:59
14 ff575c261c949d073b2895b05d1097 relationWith:11127..... 2016-02-09 16:42:59
\end{Soutput}

By default, for each artifact also it's context, i.e., session info, is saved. It can be accessed with the function \code{asession()}. See the example below. Such additional information may be very useful if we cannot replicate previous results and we are in the need of recovering the exact versions of important packages, which can be done with \code{restoreLibs} function.

\begin{Sinput}
R> asession("11127cc6ce69a89d11d0e30865a33c13")
\end{Sinput}
\begin{Soutput}
Session info ------------------------------------------------------
 setting  value                       
 version  R version 3.2.2 (2015-08-14)
 system   x86_64, darwin13.4.0        
 ui       RStudio (0.99.441)          
Packages ----------------------------------------------------------
 package      * version  date       source                         
 acepack        1.3-3.3  2013-05-03 CRAN (R 3.1.0)                 
 archivist    * 1.9.7.3  2016-02-09 CRAN (R 3.2.2)                 
 ggplot2      * 2.0.0    2015-12-16 Github (hadley/ggplot2@11679cd)
 gridExtra    * 2.0.0    2015-07-14 CRAN (R 3.2.0)                 
 ...
\end{Soutput}

\subsubsection{Serialization of an object creation event into repository}

The \pkg{archivist} provides a new operator \code{\%a\%} {that} works as the extended pipe operator \code{\%>\%} from the \pkg{magrittr} package \cite[see][for more details]{magrittr}. In addition, it saves the resulting object to the default \pkg{archivist} repository together with the function call and its parameters. The default repository should be set first, see the \code{setLocalRepo} function for instructions how to do this.
With this functionality it is possible to trace function calls and extract pedigree for some artifacts. 

\begin{Sinput}
R> library("archivist")
R> createLocalRepo("arepo", default = TRUE)
R> library("dplyr")
R> iris %a%
+    dpyr::filter(Sepal.Length < 6) %a%
+    lm(Petal.Length~Species, data=.) %a%
+    summary() -> tmp
\end{Sinput}

How to recreate an object's history? The function \code{ahistory} extracts the chain of calls that leads to the selected object. As an argument one can specify either an object's value or its MD5 hash. The value of \code{ahistory} function is a \code{data.frame} with two columns – first contains function calls while second contains MD5 hashes of partial results.

In the example above, a chain of three operations converts input \code{iris} data into the \code{tmp} object. The \pkg{dplyr} package \citep[see][]{dplyr} has to be loaded first since the function \code{filter} is used in this example. Following lines present the chain of consecutive transformations that are recorded in the repository.

\begin{Sinput}
R> ahistory(tmp)
R> ahistory(md5hash = "050e41ec3bc40b3004bc6bdd356acae7")
\end{Sinput}
\begin{Soutput}
   iris                                  [ff575c261c949d073b2895b05d1097c3]
-> filter(Sepal.Length < 6)              [d3696e13d15223c7d0bbccb33cc20a11]
-> lm(Petal.Length ~ Species, data = .)  [990861c7c27812ee959f10e5f76fe2c3]
-> summary()                             [050e41ec3bc40b3004bc6bdd356acae7]
\end{Soutput}

In order to restore an object's pedigree all partial results must be saved in a repository. So this option will work only for objects created by a chain of calls that use the \code{\%a\%} operator.

\subsubsection{Loading an object from a repository}

To read an object from repository we may consider the following four scenarios.
\begin{itemize}
\item We know the object's MD5 hash and the object is in a local directory.
\item We know the object's MD5 hash and the object is in a remote repository, i.e., on GitHub or BitBucket.
\item We do not know the hash but we know some properties of the object so we need to find it first by its tags. The object is in a local repository.
\item As above, but the object is in a remote repository.
\end{itemize}

If we know the MD5 hash of the requested artifact, we can directly load the object from the repository and in this section we are going to show how this can be done. If we do not know the MD5 hash, then we need to use one of \code{search*} functions presented in Section \ref{sec:search33}.

Functions \code{loadFromLocalRepo} and \code{loadFromRemoteRepo} read artifacts from either local or remote repositories. The local repository is defined by a path to it's root; remote repository is defined by it's type (currently \code{"github"} (default) or \code{"bitbucket"}), the username, repository's name and a subdirectory within the repository. In both functions the argument \code{value} specifies whether the function should return the object by value (\code{value=TRUE}) or it should load the object into the namespace with its original name (\code{value=FALSE}).

For the purpose of this example we have created a repository \code{graphGallery}, with two objects: a plot and a regression model. 
The repository is available both on GitHub (see \code{https://github.com/pbiecek/graphGallery}) and within the \pkg{archivist} package (see the \code{graphGallery} directory). Two archived objects have \code{7f3453331910e3f321ef97d87adb5bad} and \code{2a6e492cb6982f230e48cf46023e2e4f} hashes respectively.

The full {MD5 hash} of an artifact is a 32-characters-long string but it is enough to set only the first few characters. In the example below it is enough to use \code{"7f34533"} prefix to load an artifact with the \code{"7f3453331910e3f321ef97d87adb5bad"} hash. There is only one artifact with prefix \code{"7f34533"} in its MD5 hash. If there is more, all that match the prefix are returned. Note that one should not use this feature unless is sure that new objects with colliding hashes will not be added. For small repositories conflicts are unlikely even for first five characters, but be careful when using this feature.

\begin{figure}[h!]
\centering
\includegraphics[width=0.7\textwidth]{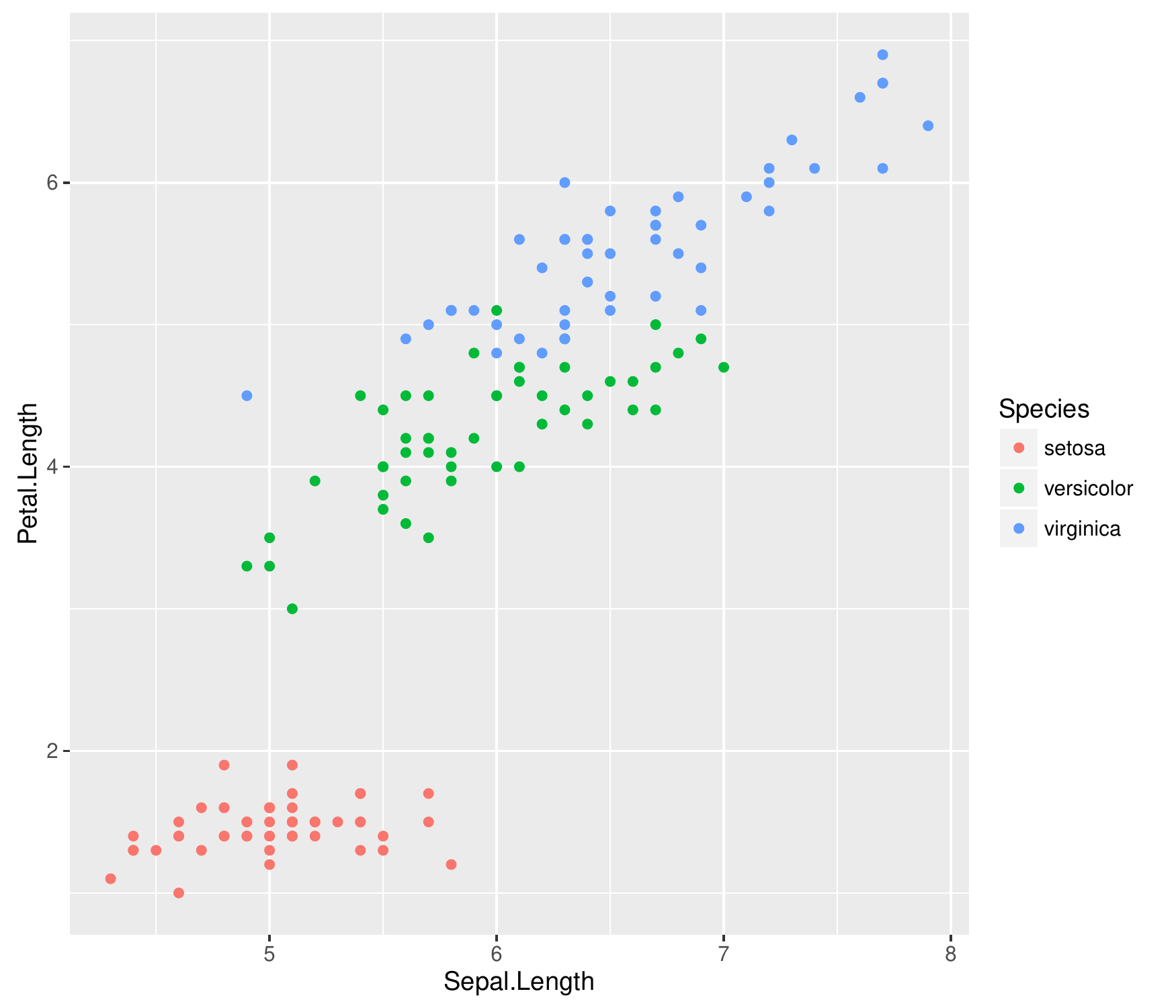}
\caption{\label{figExample1}Object with the hash  \code{7f3453331910e3f321ef97d87adb5bad} available in the repository called \code{graphGallery} of the GitHub user \code{pbiecek} and in the \proglang{archivist} package. It may be retrieved with the \code{aread} function.}
\end{figure}

Both following instructions retrieve an \proglang{R} object from GitHub, load it into  \proglang{R} session and make it accessible for further processing. In this case it is a \code{ggplot2} object so after being loaded the \code{print} function is triggered and a plot is generated  (see Figure \ref{figExample1}). Note that by default the GitHub is assumed, but this may be changed with the parameter \code{repoType}.

\begin{Sinput}
R> loadFromRemoteRepo("7f3453331910e3f321ef97d87adb5bad",
+    repo = "graphGallery", user = "pbiecek", value = TRUE)
R> loadFromLocalRepo("7f34533", 
+    system.file("graphGallery", package = "archivist"), value = TRUE)
\end{Sinput}

The \code{aread} function is a wrapper over \code{loadFromRemoteRepo} with more compact form. Shorter instructions and shorter code snippets might be placed in a figure or table caption. The single line below reads an object with the \code{7f34533...} hash from \code{graphGallery} GitHub repository  that is owned by the \code{pbiecek} user.

\begin{Sinput}
R> archivist::aread("pbiecek/graphGallery/7f3453331910e3f321ef97d87adb5bad")
\end{Sinput}

The following instructions retrieve the same \proglang{R} object but this time from the \code{graphGallery} repository attached to the \pkg{archivist} package. Note that the default repository is set first with the \code{setLocalRepo} function.

\begin{Sinput}
R> library("archivist")
R> setLocalRepo(system.file("graphGallery", package = "archivist"))
R> aread("7f3453331910e3f321ef97d87adb5bad")
\end{Sinput}

The use of MD5 hashes as objects identifiers has some advantages. 
In some use cases we may be restricted to use only models approved by some authority. For example, due to some hypothetical regulatory issues in production it might be advisable to use only a specific version of a model (such as credit scoring model or some forecasting model).

In the \pkg{archivist} package all objects have their cryptographical hash calculated with the MD5 algorithm. One can use the \code{digest} function to validate the object's MD5 hash at any moment. One can also call an object the from repository by its MD5 hash. Having a list of MD5 hashes of \textit{allowed} objects one can validate their identity.

In the example below the downloaded regression model is digested to confirm its identity.

\begin{Sinput}
R> setLocalRepo(system.file("graphGallery", package = "archivist"))
R> model <- aread("2a6e492cb6982f230e48cf46023e2e4f")
R> digest::digest(model)
\end{Sinput}
\begin{Soutput}
"2a6e492cb6982f230e48cf46023e2e4f"
\end{Soutput}

\subsubsection{Removal of an object from a repository}

To remove an artifact from a repository one can use the \code{rmFromLocalRepo} function.

In the example below the artifact \code{92ada1e052d4d963e5787bfc9c4b506c} and all its tags are removed from the repository called \code{repo}.

\begin{Sinput}
R> rmFromLocalRepo("7f3453331910e3f321ef97d87adb5bad", repoDir = repo)
\end{Sinput}

A list of artifact's hashes that should be removed may be obtained with the \code{search*} function. The example below searches for all artifacts older than 30 days and removes them from the \code{repo} repository.

\begin{Sinput}
R> obj2rm <- searchInLocalRepo(list(dateFrom = "2010-01-01", 
+    dateTo = Sys.Date()-30), repoDir = repo)
R> rmFromLocalRepo(obj2rm, repoDir = repo, many = TRUE)
\end{Sinput}

It is also possible to remove many artifacts with one call. Broader examples of this function are explained in the package manual page accessed from \proglang{R} with \code{?rmFromLocalRepo}.

\subsection{Search for an artifact and explore the repository}
\label{sec:search33}

One of the advantages of the \pkg{archivist} package is the automated derivation of artifact’s tags and meta-data. It is useful when one wants to find previously calculated results in a large collection of \proglang{R} objects. Relations between artifacts are useful when we want to process the structure dependencies between artifacts.
Below we present a list of functions for searching for artifacts on the basis of their properties.

\subsubsection{Search in a local or remote repository}

If we do not know the MD5 hashes of artifacts that are of our interest, we can find them with the use of \code{search*} functions.

Searching within a local repository and a remote repository is very similar. Functions \code{searchInLocalRepo} or \code{searchInRemoteRepo} differ only in the way in which the repository is specified.

In both functions the \code{pattern} argument may be either a tag (name, class, varname or other) or a date period in which given artifact was created. Hashes of all artifacts that meet all criteria (i.e., were created within a given time interval or have a given tag attached) are returned.

For example, the following command retrieves MD5 hashes of all objects of the class \code{gg} from the \code{pbiecek/graphGallery} repository.

\begin{Sinput}
R> searchInLocalRepo(pattern = "class:gg", 
+    repoDir = system.file("graphGallery", package = "archivist"))
\end{Sinput}
\begin{Soutput}
[1] "7f3453331910e3f321ef97d87adb5bad" "369227e67f9164dcbe934dadf2b53cc2"
\end{Soutput}

To get a list of artifacts created within a given date range one can use following instruction.

\begin{Sinput}
R> searchInLocalRepo(pattern = list(dateFrom = "2016-01-01",
+    dateTo = "2016-02-07"), 
+    repoDir = system.file("graphGallery", package = "archivist"))
\end{Sinput}
\begin{Soutput}
[1] "d9313a0de3e2980201a8971e3384ff26" "ff575c261c949d073b2895b05d1097c3"
[3] "2a6e492cb6982f230e48cf46023e2e4f" "93ecfdf1436932e2860c6dbdf2abc2ad"
[5] "afb2550d0f886f0cf3b050f04c5cd4f8"
\end{Soutput}

The \code{searchInLocalRepo} and \code{searchInRemoteRepo} functions allow to use more than one searching criteria. Additional argument \code{intersect} specifies if the resulting objects have to met all or any of the search criteria.

\begin{Sinput}
R> searchInLocalRepo(pattern=c("class:gg", "labelx:Sepal.Length"),
+    repoDir = system.file("graphGallery", package = "archivist"))
\end{Sinput}
\begin{Soutput}
[1] "369227e67f9164dcbe934dadf2b53cc2" "7f3453331910e3f321ef97d87adb5bad"
\end{Soutput}

These two functions return MD5 hashes of artifacts. In order to load these artifacts from repository one needs to use either \code{loadFrom*Repo} or \code{aread} functions.
Since both operations are usually performed together (search for MD5 hashes of artifacts by their tag / load artifacts with given MD5 hashes), one can use the \code{asearch} function which retrieves MD5 hashes and returns a list with values of artifacts that meet all selected criteria.


\subsubsection[Retrieval of a list of R objects with given tags]{Retrieval of a list of \proglang{R} objects with given tags}

When working in a team or for a longer period of time, one produces a lot of partial results and it becomes harder and harder to trace what kind of analyses were conducted in the past and where are the results. 

The \pkg{archivist} extracts meta-data from \proglang{R} objects in the very same moment they are archived in a repository. For many researchers objects are so valuable, due to their pedigree and meta-data, that they can be regarded as artifacts. Having such additional meta-data it is easier to search for previously generated partial results, e.g., by specifying what kind of model with which variables we are looking for.

For example, the code below retrieves all objects of the \code{lm} class with the \code{Sepal.Length} variable from within a list of dependent variables. In this repository only two artifacts (here \code{lm} models) match both conditions.

\clearpage

The following instruction searches within the default local repository.

\begin{Sinput}
R> setLocalRepo(system.file("graphGallery", package = "archivist"))
R> models <- asearch(patterns = c("class:lm", "coefname:Sepal.Length"))
\end{Sinput}

Below is the code that searches within the GitHub repository.

\begin{Sinput}
R> models <- asearch("pbiecek/graphGallery",  
+    patterns = c("class:lm", "coefname:Sepal.Length"))
R> lapply(models, coef)
\end{Sinput}
\begin{Soutput}
$`18a98048f0584469483afb65294ce3ed`
 (Intercept) Sepal.Length 
   -7.101443     1.858433 

$`2a6e492cb6982f230e48cf46023e2e4f`
      (Intercept)      Sepal.Length Speciesversicolor  Speciesvirginica 
       -1.7023422         0.6321099         2.2101378         3.0900021 
\end{Soutput}

The following instruction retrieves all artifacts of the \code{gg} class (created with the package \pkg{ggplot2}) with label \code{Sepal.Length} on the X axis. Two objects are returned as a result. They are plotted together by the \code{grid.arrange} function from \pkg{gridExtra} package \citep[see][]{gridExtra}.

\begin{Sinput}
R> plots <- asearch(patterns = c("class:gg", "labelx:Sepal.Length"))
R> length(plots)
\end{Sinput}
\begin{Soutput}
[1] 2
\end{Soutput}
\begin{Sinput}
R> library("gridExtra")
R> do.call(grid.arrange, plots)
\end{Sinput}

\begin{figure}[h!]
\centering
\includegraphics[width=0.7\textwidth]{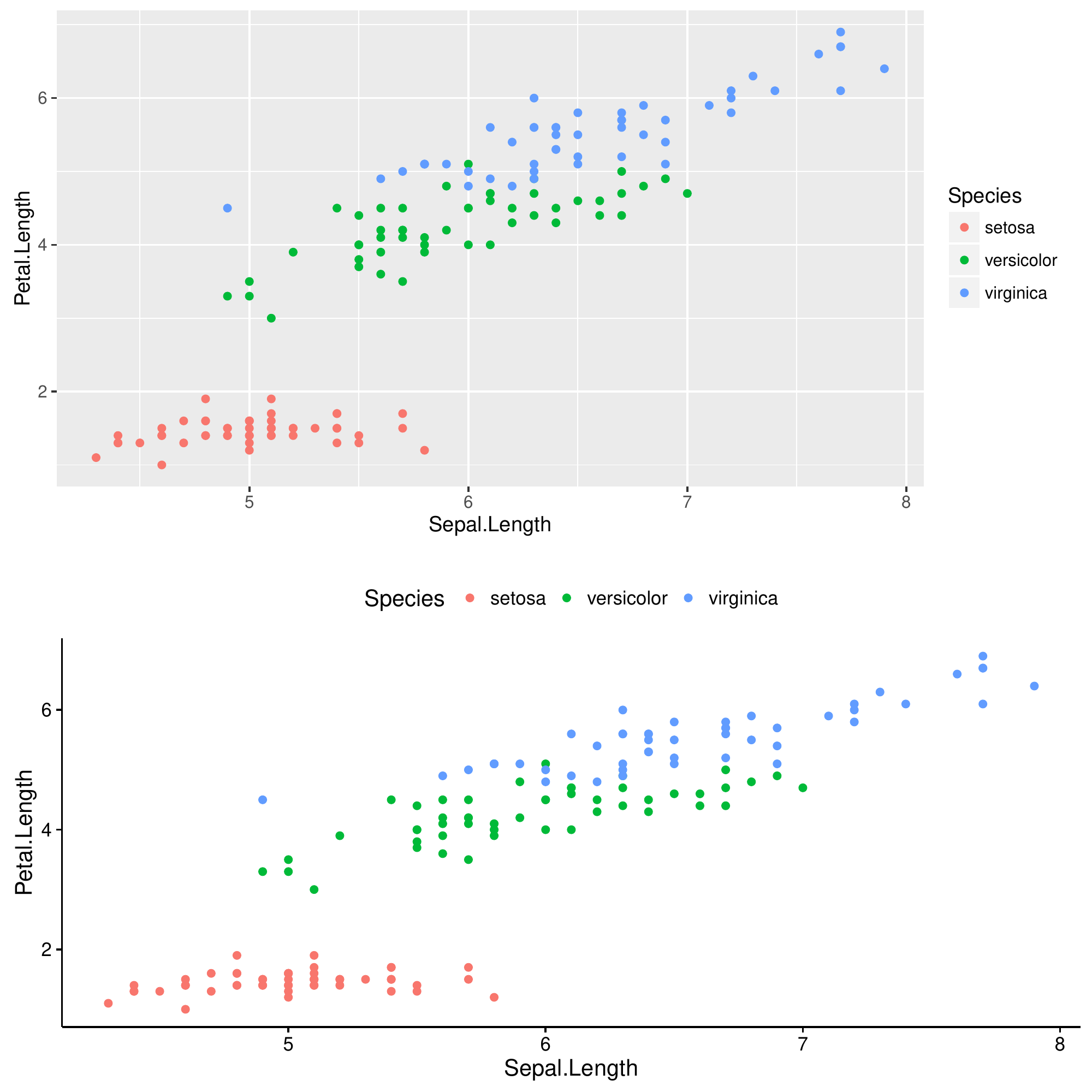}
\caption{\label{fig:twoplots}There are two objects of the class \code{gg} with annotation \code{Sepal.Length} on the {X} axis in the GitHub \code{pbiecek/graphGallery} repository. All objects in a repository that meet a set of conditions may be retrieved with the \code{asearch} function. {Instructions how to extend the list of tags are in Section \ref{sec:art32}.}}

\centering
\includegraphics[width=0.7\textwidth]{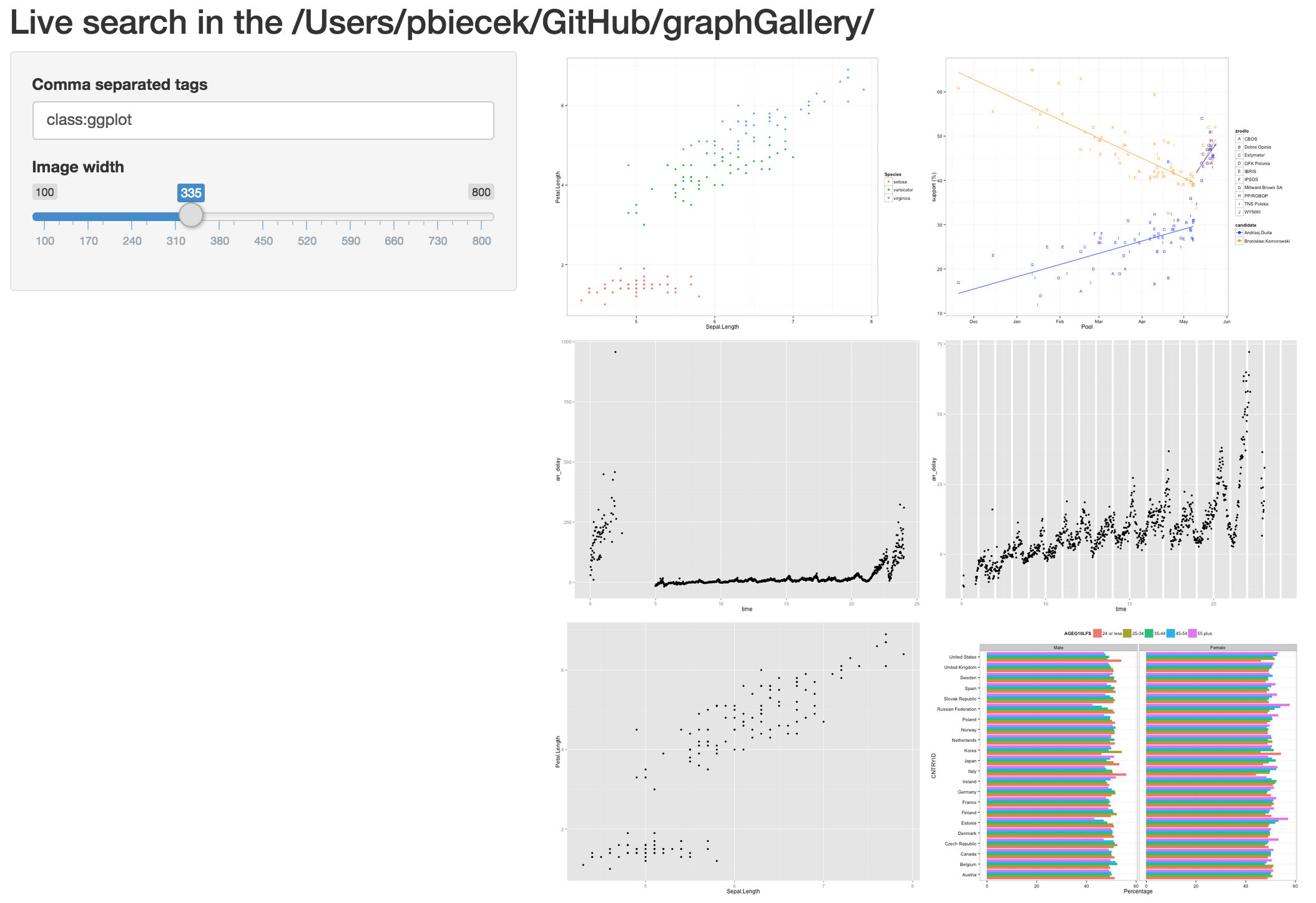}
\caption{\label{fig:shinyapp}Model screen of a Shiny application produced by \code{shinySearchInLocalRepo} function. The application helps in searching for artifacts with given tags within a selected repository.}
\end{figure}

Result of these instructions is presented in Figure \ref{fig:twoplots}.

\subsubsection{Interactive search in a local repository}

For local repositories, it is also possible to explore the repository interactively with the \code{shinySearchInLocalRepo} function. This function launches a {Shiny} application \citep[see][]{shiny} which is dynamically created and which  allows for interactive specification of tags and sorting criteria. See Figure \ref{fig:shinyapp} with an example screenshot of this application.

In the text box area one can specify tags that filter out objects presented on the right panel. Only miniatures of objects that meet all these criteria are presented. Additionally, the instruction \code{sort:key} sorts the artifacts along the key. For example, use \code{"sort:createdDate"} to sort miniatures along the date of creation of the object.

\begin{Sinput}
R> arepo <- system.file("graphGallery", package = "archivist")
R> shinySearchInLocalRepo(arepo)
\end{Sinput}

\clearpage

\subsection{Extensions}
\label{sec:extensions}

The \pkg{archivist} package is designed as a multi-purpose manager of objects. In this section we present some specific extensions.

\subsubsection{Archiving all results of a specific function}

The \code{trace()} function from the \pkg{base} package allows to insert a specific instruction to the body of a selected function. It can be used for example to call \code{saveToLocalRepo()} function at the end of a selected function.

In the example below we modify the \code{lm()} function so that after it's each execution the created \code{lm} model is automatically added to the default local repository \code{allModels}.

\begin{Sinput}
R> library("archivist")
R> createLocalRepo("allModels", default = TRUE)
R> atrace("lm", "z")
Tracing function "lm" in package "stats"
R> lm(Sepal.Length ~ Sepal.Width, data=iris)
Tracing lm(Sepal.Length ~ Sepal.Width, data = iris) on exit 

Call:
lm(formula = Sepal.Length ~ Sepal.Width, data = iris)

Coefficients:
(Intercept)  Sepal.Width  
     6.5262      -0.2234  

R> sapply(asearch("class:lm"), BIC)
42fcf77af2c40f70c445cbba513aeabd 
                        381.0236 
\end{Sinput}

\subsubsection[Integration with the knitr package]{Integration with the \pkg{knitr} package}

The \pkg{knitr} package is a tool that transforms a mixture of \proglang{R} code and descriptions in natural language into a md, html or pdf report. Moreover the produced report contains results generated by the included \proglang{R} code. On one hand reader knows that presented results are generated by presented code. On the second hand the author does not waste time on coping the results, since they are automatically included in the output. Results included in a report are usually plots or tables. In such form they cannot be loaded from the pdf/html file directly to \proglang{R}.
The \pkg{archivist} package records objects and makes them easier to access through local, GitHub or BitBucket repositories.

The function \code{addHooksToPrint} combines these two tools. A call to this function should be included on the beginning of a \pkg{knitr} report. It creates a new generic \code{print} functions for classes specified by the \code{class} argument. These functions save objects to the repository and add corresponding hooks to the report after every attempt to \code{print} the object. Hooks are short instructions on how the recorded objects can be accessed. 

An example is presented in the report \url{http://bit.ly/1nW9Cvz}. Part of this report is presented in Figure \ref{figExample3}.
On the beginning there is a snippet presented below. It automatically adds hooks to the html report for all objects of classes \code{ggplot} or \code{data.frame}. 

\begin{Sinput}
R> addHooksToPrint(class=c("ggplot", "data.frame"),
+                  repoDir = "arepo", 
+                  repo = "Eseje", user = "pbiecek", subdir = "arepo")
\end{Sinput}

As a result, just before each plot, there are automatically created hooks to corresponding objects e.g., \code{archivist::aread("pbiecek/Eseje/arepo/24ea7c04b861083d4bf56eee1c5a17b7")}. These hooks serve also as links to the corresponding \proglang{R} objects.

The biggest advantage of this integration is that a single call to \code{addHooksToPrint} is needed to enrich the \pkg{knitr} report in \pkg{archivist} hooks for all interesting objects.

\subsubsection{Gallery of artifacts in the repository}

 Information about artifacts is stored in an SQLite database in the \code{backpack.db} file. The \code{createMDGallery} function creates a single markdown file with gallery of all artifacts in the repository. 

\begin{figure}[h!]
\centering
\includegraphics[width=0.8\textwidth]{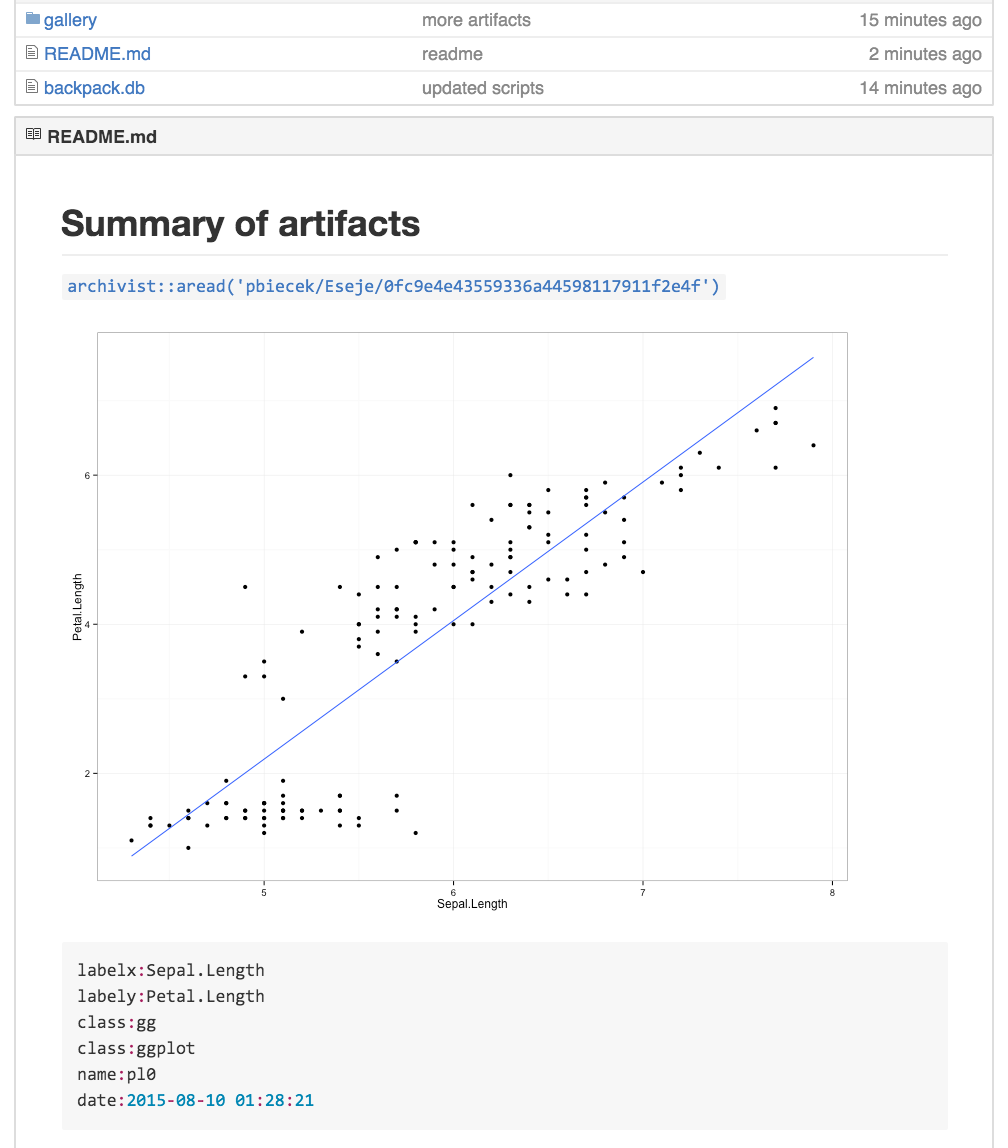}
\caption{\label{figExample4}A part of the gallery \texttt{http://bit.ly/1Q62Tpz} created with function \code{createMDGallery}. The gallery presents hooks, miniatures and list of tags for each artifact in the repository.}
\end{figure}

Such gallery, if saved as file named \code{readme.md}, will automatically list all artifacts with miniatures and tags in the GitHub web portal user interface. See an example gallery at \url{http://bit.ly/1Q62Tpz}.
This gallery was created with the following instruction. A part of the result is presented in Figure \ref{figExample4}.

\begin{Sinput}
R> createMDGallery("arepo/readme.md", 
+           repo = "Eseje", user = "pbiecek", subdir = "arepo", 
+           addMiniature = TRUE, addTags = TRUE)
\end{Sinput}

\subsubsection{Support for other repositories, other languages and other formats}

The current implementation of \pkg{archivist} supports local, GitHub and BitBucket repositories. The package is implemented in \proglang{R} and saves artifacts in the \code{rda} format. 

In order to support other repositories one can extend the function \code{getRemoteHook}. It is used internally by other \pkg{archivist} functions to generate URL addresses to files in remote repositories. In order to support other repositories it's enough to extend this function.

All metadata related to artifacts is sorted in an SQLite database in \code{backpack.db} file. This database can be accessed from other languages. Objects are stored as files and can be added in different formats. Each artifact has an additional tag \code{format:xxx} that specifies in which format the artifact is saved, one artifact can be saved in more than one format. Currently artifacts are stored as \code{rda} files. In order to save objects in other formats, like \code{json} or \code{csv}, it is enough to extend the \code{saveToLocalRepo} function. In order to load objects from other formats it is enough to overload \code{loadFromLocalRepo} and \code{loadFromRemoteRepo} functions.

\subsubsection{Restoring older versions of packages}

In some cases, in order to use an artifact it is not enough to restore it. A good example of this problem are objects of the \code{gg} class created with \pkg{ggplot2} package. The structure of \code{gg} objects is different in package \pkg{ggplot2} in the version 1.0, different in the version 2.0 and different in the version 2.1. It means that even if we have restored an object that was created with package in version 2.0 we will not be able to use the \code{plot} function for this object if one uses \pkg{ggplot2} package in the version 2.1 nor 1.0.

To use the object we need to downgrade \pkg{ggplot2} package to the version 2.0. This is possible with the \code{restoreLibs} function. For a given hash of an artifact the \code{restoreLibs} function restores it's \code{session_info} and reinstalls required packages with versions attached during the artifact's archiving. Packages can be reinstalled in the new directory, not to affect the default \proglang{R} libraries.

For example, the \code{600bda83cb840947976bd1ce3a11879d} object was created with \pkg{ggplot2} version 2.0. The \code{asession()} function checks versions of packages that were then attached.

\begin{Sinput}
R> asession("pbiecek/graphGallery/arepo/600bda83cb840947976bd1ce3a11879d")
\end{Sinput}
\begin{Soutput}
...
 Formula        1.2-1    2015-04-07 CRAN (R 3.1.3)                 
 ggplot2        2.0.0    2015-12-16 Github (hadley/ggplot2@11679cd)
 gridExtra    * 2.0.0    2015-07-14 CRAN (R 3.2.0)  
...
\end{Soutput}

Here the \pkg{ggplot2} was in the version 2.0 and was installed from GitHub. The \code{restoreLibs()} function reinstalls all libraries from proper repositories (here GitHub) to proper versions (here commit \code{11679cd}).

\begin{Sinput}
R> restoreLibs("pbiecek/graphGallery/arepo/600bda83cb840947976bd1ce3a11879d")
\end{Sinput}

After that one can load and plot the \code{ggplot} object since the structure of \code{gg} object is compatible with installed libraries.

\begin{Sinput}
R> aread("pbiecek/graphGallery/arepo/600bda83cb840947976bd1ce3a11879d")
\end{Sinput}

\section{Conclusions}
\label{sec:conc4}

The goal of a data analysis is not only to answer a research question based on data but also to collect findings that support that answer. These findings usually take the form of a~table, plot or regression/classification model and are usually presented in articles or reports. Such objects are mostly well presented graphically, but they are hard to recreate back in a~computer.

In this paper we have presented the \proglang{R} package called \pkg{archivist}, which implements the logic of recordable research. The \pkg{archivist} stores \proglang{R} objects in repositories. The data scientist may share obtained results with other users, create hooks to models and then embed these hooks in articles, reports or web applications. One may also search within a repository and look for artifacts with given properties or relations with other artifacts. One may also validate the object’s identity or derive its pedigree.

Repositories may be shared among team members or between different computers or systems. Statistical models or plots may be stored in a single repository which simplifies the object management.

In this article we have also presented some use-cases for the \pkg{archivist} package, such as: hooks for \proglang{R} objects that can be embedded in reports or articles, interactive searching within repository or retrieving object’s pedigree.

\section{Acknowledgments}

Thanks go to Ross Ihaka, \L{}ukasz Bartnik, Cezary Chudzian and two anonymous reviewers for valuable discussions and comments on the idea of recordable research and early versions of this paper. We would like to thank Witold Chodor for his great contributions to the development of this package. The package \pkg{archivist} was initiated as an open project in the company \textit{iQor Polska sp. z o.o.}.

\bibliography{jss2342}

\end{document}